\documentclass[aps,prx,reprint,showpacs,nosuperscriptaddress,floatfix]{revtex4-2}
\usepackage[utf8]{inputenc}
\usepackage{graphicx}

\usepackage{hyperref}
\hypersetup{
colorlinks=true,final=true,
        linkcolor=blue,
        citecolor=blue,
        filecolor=blue,
        urlcolor=blue,
}
\usepackage{color}
\usepackage{ulem}

\begin{document}

\title{Electronic structure of Ruddlesden-Popper nickelates: Strain to mimic the effects pressure}

\author{Yi-Feng Zhao}
 \altaffiliation {yzhao421@asu.edu}
\author{Antia S. Botana}%
\affiliation{%
Department of Physics, Arizona State University, Tempe, AZ 85287, USA}%

\date{\today}

\begin{abstract}
Signatures of superconductivity under pressure have recently been reported in the bilayer La$_3$Ni$_2$O$_7$ and trilayer La$_4$Ni$_3$O$_{10}$ Ruddlesden-Popper (RP) nickelates with general chemical formula La$_{n+1}$Ni$_n$O$_{3n+1}$ ($n=$ number of perovskite layers along the $c$-axis). The emergence of superconductivity is always concomitant with a structural transition in which the octahedral tilts are suppressed, bringing the apical Ni-O-Ni angle to 180$^\circ$ and causing an increase in the out-of-plane $d_{z^2}$ orbital overlap. Due to this strong interlayer coupling, a flat band of pure $d_{z^2}$ character crosses the Fermi level. Here, using first-principles calculations, we explore biaxial strain (both compressive and tensile) as a means to mimic the electronic structure characteristics of RP nickelates (up to $n=5$) under hydrostatic pressure. Our findings highlight that strain allows to decouple the structural and electronic structure effects obtained under hydrostatic pressure: while compressive strain brings the apical Ni-O-Ni angle closer to 180$^\circ$, it shifts the $d_{z^2}$ flat bands away from the Fermi energy, giving rise to a more cuprate-like electronic structure. In contrast, tensile strain reduces the apical Ni-O-Ni angle (to values $\sim$ 160$^\circ$), but it recovers the flat $d_{z^2}$ band at the Fermi level appearing in the bilayer and trilayer RPs under pressure. Overall, strain represents a promising way to tune the electronic structure of RP nickelates and could be an alternative route to achieve superconductivity at ambient pressure in this family of materials.

\end{abstract}

\maketitle


\section{\label{sec:level1} INTRODUCTION}

The beginning of the nickel age of superconductivity was marked by the report of a superconducting transition in thin films of Sr-doped infinite-layer NdNiO$_2$ (with $T_c$ $\sim$ 15 K) in 2019~\cite{li2019superconductivity}. Follow-up studies showed that the Pr- and La- infinite-layer analogs are also superconducting~\cite{osada2020superconducting, Osada2020, osada2021nickelate, zeng2022superconductivity, dome_nd}. Subsequently, superconductivity was also found in films of the quintuple-layer compound Nd$_6$Ni$_5$O$_{12}$ with a similar $T_c$~\cite{pan2022superconductivity}. These materials are members of the structural series with general chemical formula $R_{n+1}$Ni$_{n}$O$_{2n+2}$ ($R$ = rare-earth), characterized by a sequence of $n$-NiO$_{2}$ layers along the $c$-axis and a cuprate-like electron filling close to $d^9$ \cite{Lacorre1992, Poltavets2007crystal, Polatvets2006la326, labollita2021reduced}. A body of work has suggested that a model based on a single Ni-$d_{x^2-y^2}$ correlated band, with
additional weakly correlated $R$-$d$ bands that act as a charge reservoir is appropriate to describe the electronic structure of these materials~\cite{kitatani2020nickelate,botana2020similarities,saka2020,Worm2022.6.L091801, Karp2020manybody, Karp2020comparative}. In terms of pairing, most evidence, both theoretical and experimental, points towards a dominant $d$-wave instability~\cite{Wu2020,Kitatani2023,cheng2024evidence}.  In spite of their obvious similarities to the cuprates, the superconducting layered nickelates show some relevant differences, the most noticeable one being their lower $T_c$. Further, no bulk superconductivity in this family has been reported to date \cite{wang2020absence, Li2020absence}.

Recently, signatures of superconductivity with much higher $T_c$ were found in bulk samples of the bilayer ($n=2$) and trilayer ($n=3$)  Ruddlesden-Popper (RP) $R_{n+1}$Ni$_{n}$O$_{3n+1}$ nickelates under pressure~\cite{sun2023signatures,wang2024bulk, hou2023emergence, zhu2024superconductivity,li2023signature}. The $R_{n+1}$Ni$_{n}$O$_{3n+1}$ compounds represent the parent phases of the reduced layered $R_{n+1}$Ni$_{n}$O$_{2n+2}$ counterparts \cite{Greeenblatt1997ruddlesden}.  In the bilayer RP La$_3$Ni$_2$O$_7$, a $T_c$ $\sim$ 80 K has been reported at P $\sim$ $14-40$ GPa \cite{sun2023signatures,wang2024bulk, hou2023emergence}. In the trilayer
nickelate La$_4$Ni$_3$O$_{10}$ a $T_c$ $\sim$ 30 K has been found at $P\sim$ 30 GPa \cite{zhu2024superconductivity, li2023signature}. These RP materials possess 
$n$-NiO$_6$-perovskite layers and have an average $d^{7+\frac{1}{n}}$ filling, further from cuprate values~\cite{jung2022electronic}. In both La$_3$Ni$_2$O$_7$ and La$_4$Ni$_3$O$_{10}$, pressure gives rise to a structural transition, concomitant with the onset of superconductivity, associated with the suppression of octahedral tilts in the perovskite blocks \cite{sun2023signatures,hou2023emergence,wang2024bulk,zhu2024superconductivity}. In terms of their electronic structure,  one key difference with respect to the layered materials is that two $d$-orbitals (Ni-$d_{z^2}$ and $d_{x^2-y^2}$) are important in both La$_3$Ni$_2$O$_7$ and La$_4$Ni$_3$O$_{10}$ \cite{luo2023bilayer, Zhang2023, Yang2023, Yang2024, Zhang2024save, sakakibara2024, christiansson2023correlated, labollita2024electronic, rhodes2024structural}. Based on a two-orbital model, a number of theoretical studies have found that, linked to changes in the fermiology under pressure (in particular, to the emergence of a flat band of pure $d_{z^2}$ character crossing the Fermi level), the leading superconducting instability in the bilayer and trilayer nickelate RPs has $s^{\pm}$ symmetry~\cite{lechermann2023electronic, Yang2024,Zhang2024,Zhang2024save, zhang2024structural, Lu2024}.

Ultimately, experimental discovery in the families of the superconducting reduced and parent-RP nickelates has been mutually exclusive, with the former being restricted to thin films and the latter to bulk samples. Given the intricacies of applying hydrostatic pressure and the related reproducibility issues \cite{liling}, epitaxial strain in thin films could be an alternative route to drive (at ambient pressure) the structural and electronic structure characteristics associated with the onset of superconductivity in RP nickelates. It can further be a means to expand this family of superconductors: RPs with $n>3$ are not thermodynamically stable in the bulk but in thin film form RP nickelates with $n$ up to 5 have been grown~\cite{li2020epitaxial,Sun2021,Pan2022}.

Here, using first-principles calculations, we systematically explore the structure and electronic structure of strained RP nickelates La$_{n+1}$Ni$_{n}$O$_{3n+1}$ ($n= 2-5$) focusing primarily on 3\% compressive and tensile biaxial strain levels. We find that strain allows to decouple the effects that hydrostatic pressure application produces in the structure and electronic structure characteristics of RP nickelates: while a suppression of the octahedral tilts is obtained under compressive strain, the electronic structure does not resemble that derived under hydrostatic pressure as the flat $d_{z^2}$ band is shifted down in energy below the Fermi level.  In contrast, while upon tensile strain the tilting of the octahedra increases, the associated electronic structure recovers the $d_{z^2}$ band crossing the Fermi energy that has been related to the onset of superconductivity in the bilayer and trilayer RPs. Our results suggest that biaxial strain can be used to mimic the effects of hydrostatic pressure in RP nickelates and hence it might be a means to replicate the emergence of superconducting signatures at ambient pressure in thin films, as well as to interpolate between different pairing symmetries. While our primary goal is to simply obtain general trends from computational experiments, the strain levels we have used could likely be attained epitaxially using as substrates YAlO$_3$ (compressive strain) and DyScO$_3$ (tensile strain).

\section{\label{sec:level2} computational methods}

We started by performing structural relaxations for all La$_{n+1}$Ni$_{n}$O$_{3n+1}$ ($n= 2-5$) under different strain levels using density-functional theory (DFT) calculations. To that end, we constrained the in-plane lattice constants to the appropriate strain level (biaxial strain, $\epsilon=\frac{a^*(b^*)}{a(b)}$)
and relaxed both the out-of-plane lattice constant and the internal coordinates. Bilayer La$_3$Ni$_2$O$_7$  undergoes a structural transition under pressure from orthorhombic \textit{Amam} to \textit{Fmmm} at 14 $\sim$ GPa, concomitant with the emergence of superconductivity \cite{sun2023signatures}. This transition leads to a suppression of the octahedral tilts as reflected in the change of the Ni-O-Ni bond angle across the apical oxygens ($\theta$ in Fig.~\ref{fig1}(a)) from 168.0$^{\circ}$ to 180.0$^{\circ}$ \cite{sun2023signatures,zhang2024structural}. Hence, we performed a comparison between strain applied to the $Amam$  phase of  La$_3$Ni$_2$O$_7$ and the corresponding `tetragonalized' $Fmmm$ structure ~\cite{sun2023signatures}. We analyzed all strain levels between $-3$ (compressive) and $+3\%$  (tensile) in 1\% increments. For the trilayer  La$_4$Ni$_3$O$_{10}$ we used both the orthorhombic $Bmab$ as well as a tetragonal $I4/mmm$ structure as starting points \cite{Zhang2020,jung2022electronic} and simply analyzed extreme values of strain of $\pm$ 3\%.  
Given that the structures of the higher-order RP materials ($n=4-5$) have not yet been experimentally resolved, we directly used the tetragonal $I4/mmm$ structures reported in previous work as a starting point~\cite{jung2022electronic} and analyzed strains of $\pm$ 3\% as well. The structures under pressure that we report were obtained using the procedures reported in previous work \cite{labollita2024electronic, labollita2024trilayer}. 
The Vienna $ab~initio$ Simulation Package (VASP) was employed for the structural optimizations~\cite{kresse1993ab,kresse1996efficient,blochl1994projector}. As the exchange-correlation functional we used the Perdew-Burke-Ernzerhof (PBE) version of the generalized gradient approximation (GGA)~\cite{perdew1996generalized}. The energy cutoff was set to 500 eV and a $k$-point mesh $8\times8\times2$ was used in the relaxation procedure for all the  RP phases. All the ionic positions were fully optimized until the Hellman-Feynman force was lower than 1 meV/\AA. 

With the optimized strained structures, we subsequently studied the electronic structure of all the $n=2-5$ La-RP nickelates utilizing the all-electron full-potential DFT code WIEN2K~\cite{blaha2020wien2k}. The exchange-correlation functional used in our WIEN2K calculations was also GGA-PBE~\cite{perdew1996generalized}. Only non-magnetic calculations were performed for all RPs analyzed here. We used an $R_{MT}K_{\rm max}=7.0$ and muffin-tin
radii of 2.39, 1.85, and 1.64 a.u. for La, Ni, and O atoms, respectively. A fine $k$ mesh of $11\times11\times11$ points was employed for all RP nickelates. To better understand the electronic structure trends under strain, we used maximally localized Wannier functions (MLWFs) employing Wien2Wannier and Wannier90 \cite{kunevs2010wien2wannier,Pizzi2020}. 
To downfold the DFT results onto MLWFs we chose all five Ni-$d$ orbitals to fit the electronic structure. The wannier fits to the DFT bands are shown in Fig. \ref{figs1} of Appendix \ref{appendix:A}.

\section{\label{sec:level3} RESULTS}
\subsection{Bilayer RP nickelate}

\begin{figure}
    \centering
    \includegraphics[width=1.0\linewidth]{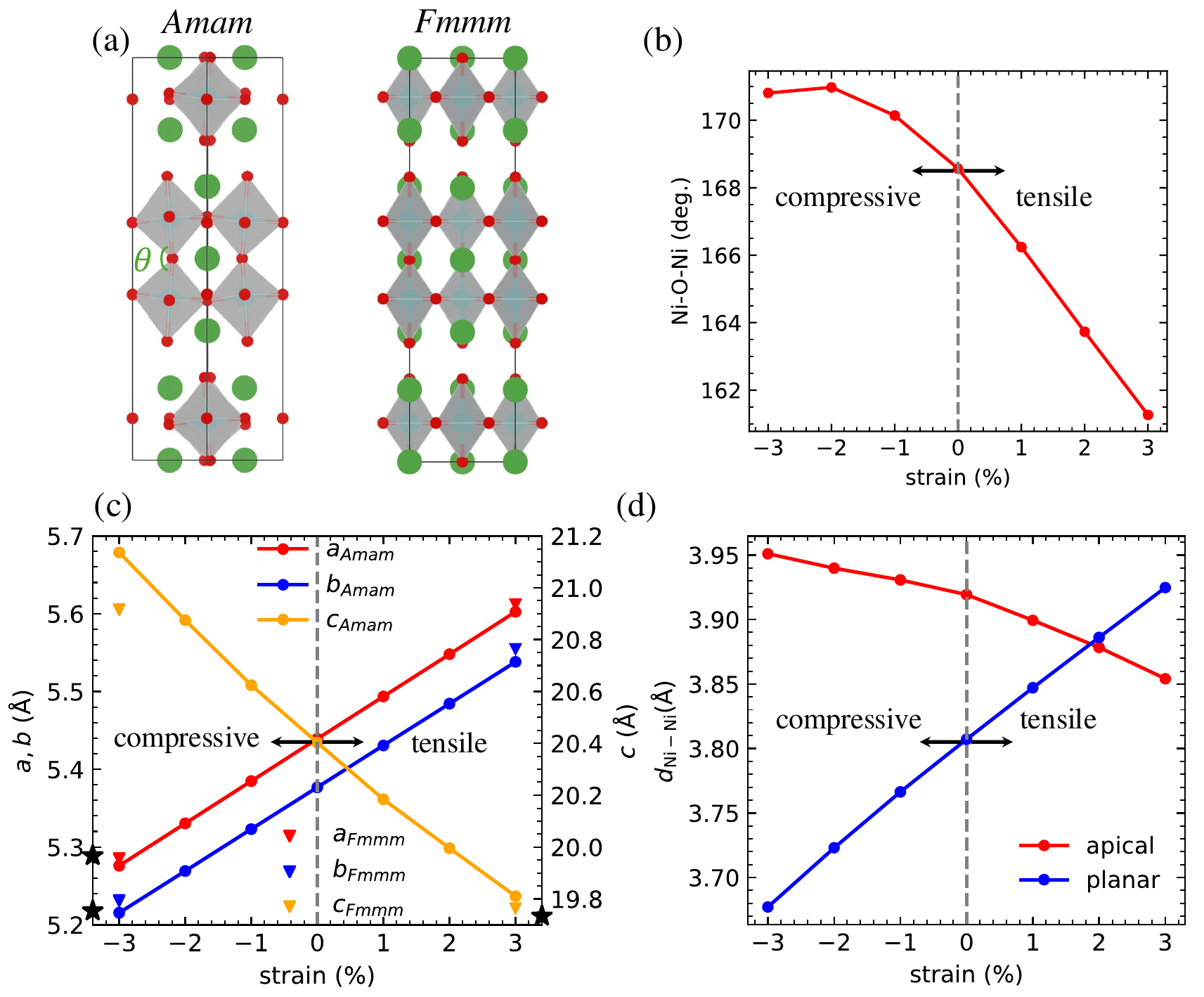}
    \caption{(a) Structure of the bilayer ($n=2$) RP La$_3$Ni$_2$O$_7$ depicted for both the $Amam$ (left panel) and $Fmmm$ (right panel). The green, and red spheres represent the lanthanum and oxygen atoms, respectively. The Ni atoms are inside the gray octahedra. (b) Change in the Ni-O-Ni bond angle across the apical oxygen ($\theta$ in panel (a)) upon strain for the $Amam$ structure. (c) Evolution of the lattice parameters of La$_3$Ni$_2$O$_7$ upon strain for the $Amam$ structure (circles) and for the $Fmmm$ structure (triangles). The lattice constants of the bulk at 30 GPa are marked on the corresponding axis as black stars for comparison. (d) Evolution of the apical and planar Ni-Ni bond lengths for the $Amam$ structure upon strain.}
    \label{fig1}
\end{figure}

We first focus on analyzing the results obtained for strained bilayer  La$_3$Ni$_2$O$_7$ which we use as a benchmark to subsequently analyze the structures and electronic structures of the nickelate RPs with higher $n$ values. We start by looking at the structural data for La$_3$Ni$_2$O$_7$.  Several studies have suggested that the straightening of the Ni-O-Ni bond angle along the $c$-axis  under pressure described above is likely related to the emergence of superconductivity in due to the associated enhancement of the interlayer coupling \cite{wang2024structure, Lu2024}. As shown in Fig. \ref{fig1}(b), we find that compressive strain (denoted with negative numbers) also tends to `tetragonalize'  the $Amam$ structure with the Ni-O-Ni bond angle becoming $\sim$~172$^{\circ}$ for a -3\% strain. In contrast, tensile strain (denoted with positive numbers) increases the octahedral tilts with an out-of-plane Ni-O-Ni bond angle of $\sim$~161$^{\circ}$ being obtained for a +3\% strain.

Figure~\ref{fig1} exhibits further structural information for strained La$_3$Ni$_2$O$_7$ that is shown in comparison to the data obtained at 30 GPa. Importantly, as shown in Fig. \ref{fig1}(c), a 3\% tensile strain allows to closely match the out-of-plane lattice constant obtained at 30 GPa ($c$ $\sim$ 19.7 \AA) while a 3\% compressive strain sharply matches the in-plane lattice constants obtained at 30 GPa ($a$ $\sim$ 5.28 \AA ~and $b$ $\sim$ 5.22 \AA). Similar conclusions can be drawn concerning both the planar and apical Ni-Ni distances, whose evolution with strain can be seen in Fig.~\ref{fig1}(d).  Note also that the structural data obtained for the $\pm$3\% strain levels is very similar in $Fmmm$ and $Amam$ symmetries, in agreement with the results obtained under pressure \cite{labollita2024electronic}. In contrast, a 1\% strain seems to put the lattice parameters outside the superconducting regime as they can be closely matched to those obtained at 5 GPa \cite{labollita2024electronic}. To summarize the structural trends we obtain for strained La$_3$Ni$_2$O$_7$: compressive strain brings the Ni-O-Ni bond angle across the apical oxygens closer to 180$^{\circ}$, it enlarges the Ni-Ni out-of-plane distance (as expected), and it reduces the Ni-Ni in-plane distance so that for a 3\% strain the latter can be closely matched to that obtained at 30 GPa. Conversely, tensile strain reduces the out-of-plane Ni-O-Ni bond angle with respect to that at ambient pressure, it gives rise to an increased Ni-Ni in-plane distance, and it reduces the Ni-Ni out-of-plane distance so that for a 3\% strain the latter can be matched to that obtained at 30 GPa. In this manner, strain should enable to decouple the effects of planar vs. out-of-plane Ni-Ni bond-length compressions, and it should also allow discerning what effects are related to the suppression of octahedral tilts in the electronic structure, as we will show below.

\begin{figure}
    \centering
    \includegraphics[width=1.0\linewidth]{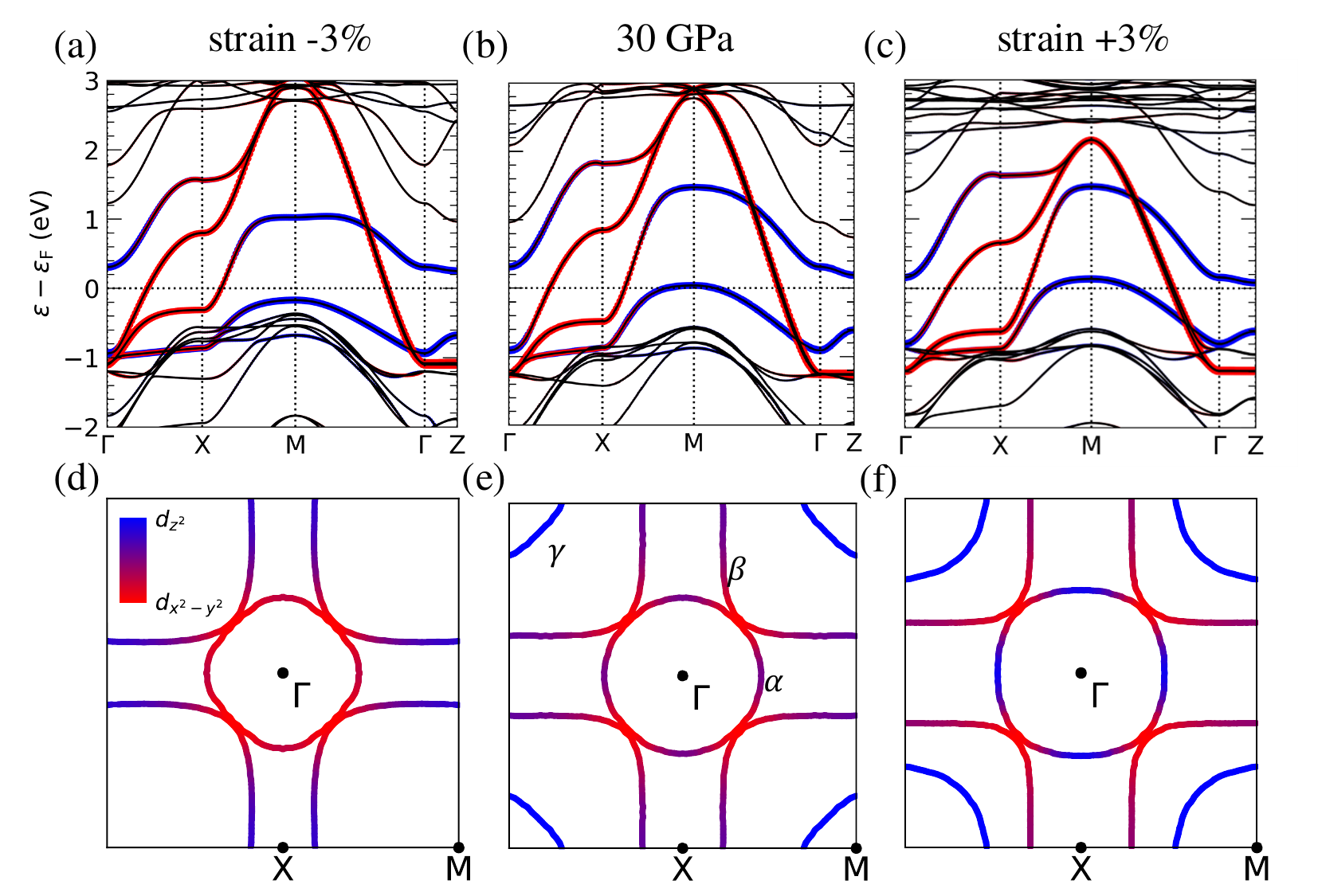}
    \caption{Band structure and Fermi surface of La$_3$Ni$_2$O$_7$ ($n=2$ RP) under different strains and under pressure with orbital characters highlighted in $Fmmm$ symmetry: (a) and (d) panels for a $-3\%$ (compressive) strain, (b) and (e) panels for 30 GPa, and (c) and (f) panels for a $+3\%$ (tensile) strain.}
    \label{fig2}
\end{figure}

In Fig. \ref{fig2} the band structures along high-symmetry directions and the corresponding Fermi surfaces are shown for La$_3$Ni$_2$O$_7$ at 30 GPa as well as for the 3\% compressive and tensile strain cases -all of them are shown in $Fmmm$ symmetry as this is the space group attained under pressure (see Fig. \ref{figs2} of Appendix \ref{appendix:E} for the corresponding band structures in $Amam$ symmetry that show the same overall trends). We start by revisiting the electronic structure at 30 GPa that has extensively been described in the literature \cite{luo2023bilayer, Zhang2023, Yang2023, Yang2024, Zhang2024save, sakakibara2024, christiansson2023correlated, lechermann2023electronic, rhodes2024structural, Yang2024,Zhang2024,Zhang2024save, zhang2024structural, Lu2024, labollita2024electronic, christiansson2023correlated}. The band structure near the Fermi level is characterized by Ni-e$_g$ ($d_{z^2}$ (blue) and $d_{x^2-y^2}$ (red)) states hybridized with O (2$p$) states (see the density of states (DOS) in Fig. \ref{figS4} of Appendix \ref{appendix:C}), indicating that these two Ni-$d$ orbitals play the dominant role in the low-energy electronic structure of this material. This is expected as nominally the Ni valence in the bilayer RP is 2.5+ (corresponding to $d^{7.5}$). This average $d$ filling means
that 1.5 e$_g$-electrons per Ni (3 per bilayer) need to be accommodated
close to the Fermi level, while the $t_{2g}$ orbitals are filled. The Ni-$d_{z^2}$ states are split by $\sim$ 1 eV into a bonding and antibonding molecular-orbital combination due to the quantum confinement
of the nickel-oxygen bilayers in the structure. As mentioned above, the Ni-$d_{z^2}$ bonding band crosses the Fermi level under pressure at M. The Ni-$d_{x^2-y^2}$ dispersion is large with a bandwidth $\sim4$ eV and this orbital remains only partially occupied. 

The accompanying Fermi surface at 30 GPa is characterized by one electron-like pocket around $\Gamma$ ($\alpha$) and one hole-like sheet opening towards X ($\beta$), both of mixed $d_{z^2}$ and $d_{x^2-y^2}$ orbital characters. The extra pocket of pure $d_{z^2}$ character ($\gamma$) around M is associated with the flat $d_{z^2}$ bonding band. This pure $d_{z^2}$ pocket is the most important change with respect to the fermiology at ambient pressure wherein the $\gamma$ band seems to be absent, as shown by ARPES \cite{Yang2024arpes}. The presence of this extra pocket also makes the electronic structure of pressurized La$_3$Ni$_2$O$_7$ markedly different from that of the cuprates.

The corresponding band structures under strain near the Fermi level are still characterized by the presence of Ni bands with dominant $e_g$ orbital character but some important differences arise: under compressive strain ($-3\%$) the $d_{x^2-y^2}$ bandwidth is comparable to that obtained at 30 GPa (as a consequence of the in-plane compression) but the bonding $d_{z^2}$ band around M is now shifted down in energy and fully occupied. As such, the $d_{z^2}$-$\gamma$ pocket is consequently absent from the Fermi surface that only preserves the $\alpha$ and $\beta$ sheets of mixed orbital character resembling the electronic structure of bilayer cuprates even though with clear $d_{z^2}$ admixture. In contrast, under tensile strain (+3\%), in spite of the very much reduced $d_{x^2-y^2}$ bandwidth, the flat $d_{z^2}$ bonding band now crosses the Fermi level giving rise to the $\gamma$ pocket. For a 3\% tensile strain, this pocket is larger in size than that obtained at 30 GPa but we are interested here in showing trends rather than on precisely matching the pressurized electronic structure. The analogous plots for a $\pm$1\% strain are shown in Fig.~\ref{figs3} of Appendix \ref{appendix:B}, where it can be seen that for La$_3$Ni$_2$O$_7$, a 1\% tensile strain actually seems to reproduce the electronic structure at 30 GPa better. However, as the lattice constants obtained for a 1\% strain are outside of the superconducting regime in La$_3$Ni$_2$O$_7$ (as described above), and as for La$_4$Ni$_3$O$_{10}$ higher pressures are needed to achieve superconductivity, we focus on the main text on larger levels of strain that can more clearly emphasize trends in the electronic structure. In that sense, it is clear that if one wanted to mimic the electronic structure of La$_3$Ni$_2$O$_7$ under hydrostatic pressure with strain, the tensile route is probably the appropriate one as it gives rise to the extra $\gamma$ pocket, in contrast to compressive strain that suppresses it further. This `desired' $\gamma$-pocket feature in the electronic structure arises even though of one of the structural effects under tensile strain is `undesired', namely, the increase in the out-of-plane Ni-O-Ni bond angle. Given that strain nicely enables this decoupling between structural and electronic structure effects, we conclude that the compression along the $c$ axis (regardless of the octahedral tilt angle) seems to be the most crucial parameter to tune the electronic structure to ultimately mimic the effects of pressure. Regardless, the compressive strain route is also an interesting one to pursue as it gives rise to a cuprate-like band structure and a larger bandwidth.   

We note that the appearance under pressure of the additional $\gamma$ pocket of pure $d_{z^2}$ character in the Fermi surface has been associated with the emergence of superconductivity in the literature. Several calculations analyzing superconducting pairing mechanisms in La$_3$Ni$_2$O$_7$ (based on a bilayer two-orbital model) find a leading $s^{\pm}$ instability under pressure that is induced precisely by the presence of this pocket \cite{lechermann2023electronic, Yang2024,Zhang2024,Zhang2024save, zhang2024structural, Lu2024}. If the $\gamma$ pocket disappears (as it happens at ambient pressure but also under compressive strain, as shown above), some works have pointed out the possibility of $d$-wave pairing symmetry instead \cite{zhang2024structural}. This possibility makes compressive strain a very compelling direction to scrutinize.  Overall, the presence of different orbitals may lead to competition between different pairing symmetries \cite{lechermann2023electronic, hanghui2024electronic, Lu2024} and we speculate that strain could perhaps also enable their decoupling. Given the above description of the electronic structure, strain could actually shed light on whether the presence of the $\gamma$ pocket is actually necessary for superconductivity to arise in La$_3$Ni$_2$O$_7$ (as it is absent under compressive strain). While it would be interesting to explicitly analyze the trends in pairing symmetry when transitioning between the compressive and tensile strain regimes in RPs, we leave that for future work. 

\begin{figure}
    \centering
    \includegraphics[width=1.0\linewidth]{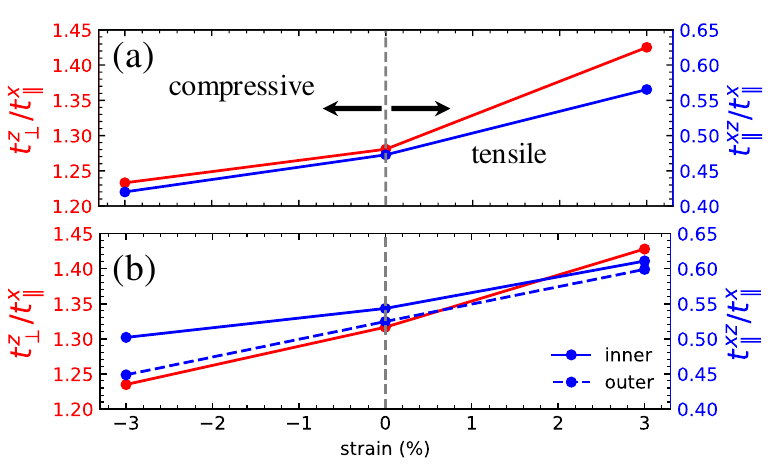}
    \caption{Relevant hopping ratios as a function of strain for (a) La$_3$Ni$_2$O$_7$ and (b) La$_4$Ni$_3$O$_{10}$ extracted from MLWFs. $t_\perp^z$ represents the out-of-plane hopping between $d_{z^2}$ orbitals,  $t_{\parallel}^x$ in-plane between $d_{x^2-y^2}$ orbitals, and $t_{\parallel}^{xz}$ in-plane between $d_{x^2-y^2}$ and  $d_{z^2}$ orbitals.}
    \label{fig3}
\end{figure}

To give further insights into the effects of strain in the electronic structure of La$_3$Ni$_2$O$_7$, we also analyze the evolution of the dominant hoppings obtained from the wannierizations. As it has been shown in previous work, there is a strong interlayer coupling within the bilayer in La$_3$Ni$_2$O$_7$ as a consequence of the overlap between the $d_{z^2}$ orbitals via the O-$p_z$ orbitals reflected in a large $t_\perp^z$ hopping $\sim$ 0.6 eV at 30 GPa. There is also a relatively large nearest-neighbor hopping integral for the $d_{x^2-y^2}$ orbitals $t_{\parallel}^x$ $\sim$ 0.4 eV as well as a sizable degree of hybridization in-plane between $d_{x^2-y^2}$ and  $d_{z^2}$ orbitals $t_{\parallel}^{xz}$ $\sim$ 0.2 eV. All of these hopping values are significantly increased with respect to their ambient pressure counterparts. It has been shown that the associated ratio of $J_{\perp}$/$J_{\parallel}$ (proportional to $t_{\perp}^z$/$t_{\parallel}^x$) could change the pairing symmetry from $s$-wave to $d$-wave, with the former being favored for the regime of larger ratios of $J_{\perp}$/$J_{\parallel}$ obtained under pressure \cite{Lu2024}. In Fig. \ref{fig3}(a) we show the evolution of the dominant hopping ratios under strain for La$_3$Ni$_2$O$_7$ where it can be seen that the $t_{\perp}^z$/$t_{\parallel}^x$ ratio increases under tensile strain and decreases under compressive strain, further reinforcing the argument exposed above that strain could likely serve as a means to tune the pairing symmetry if superconductivity can be achieved. It can also be seen how the degree of hybridization between $d_{x^2-y^2}$ and  $d_{z^2}$ orbitals increases considerably when a tensile strain is applied, as it happens with pressure.

\subsection{Trilayer RP nickelate}

Signatures of superconductivity under pressure have also been reported in the trilayer nickelate La$_4$Ni$_3$O$_{10}$ under pressure (albeit with lower T$_c$ $\sim$ 30 K and at higher pressures than those required in the bilayer material) \cite{zhu2024superconductivity}. We now study the structural and electronic structure trends obtained for La$_4$Ni$_3$O$_{10}$ focusing on a 3\% compressive and tensile strain and analyze if strain can also mimic the effects of hydrostatic pressure in this system. 

In terms of the structure, similar trends to those described above for the bilayer nickelate are obtained. For La$_4$Ni$_3$O$_{10}$ under pressure, the structure also undergoes a tetragonalization at $\sim$ 14 GPa through which the octahedral tilts are suppressed and the space group symmetry changes from $P2_1/a$ to $I4/mmm$ \cite{zhu2024superconductivity} (from first principles calculations a similar transition can be obtained starting from either the $Bmab$ or $P2_{1}/a$ structures at ambient pressure \cite{labollita2024trilayer}). Here, to analyze the effects of strain (restricted to $\pm$3\%), we start with the experimental structure resolved in $Bmab$ symmetry \cite{Zhang2020} wherein the Ni-O-Ni angle across the apical oxygens is $\sim$ 165$^{\circ}$. Under a 3\% compressive strain, the octahedral tilts are considerably suppressed with the angle transforming to $\sim$ 170$^{\circ}$ whereas under the same tensile strain, the out-of-plane Ni-O-Ni angle is reduced becoming $\sim$ 156$^{\circ}$. The lattice constants at 30 GPa ($a_{\rm{tetragonal}}=$ 3.70 \AA~ and $c=$ 26.71 \AA) can be closely matched to those obtained for a 3\% compressive ($a=$ 3.73 \AA) and tensile strain ($c=$ 27.26 \AA).

\begin{figure}
    \centering
    \includegraphics[width=1.0\linewidth]{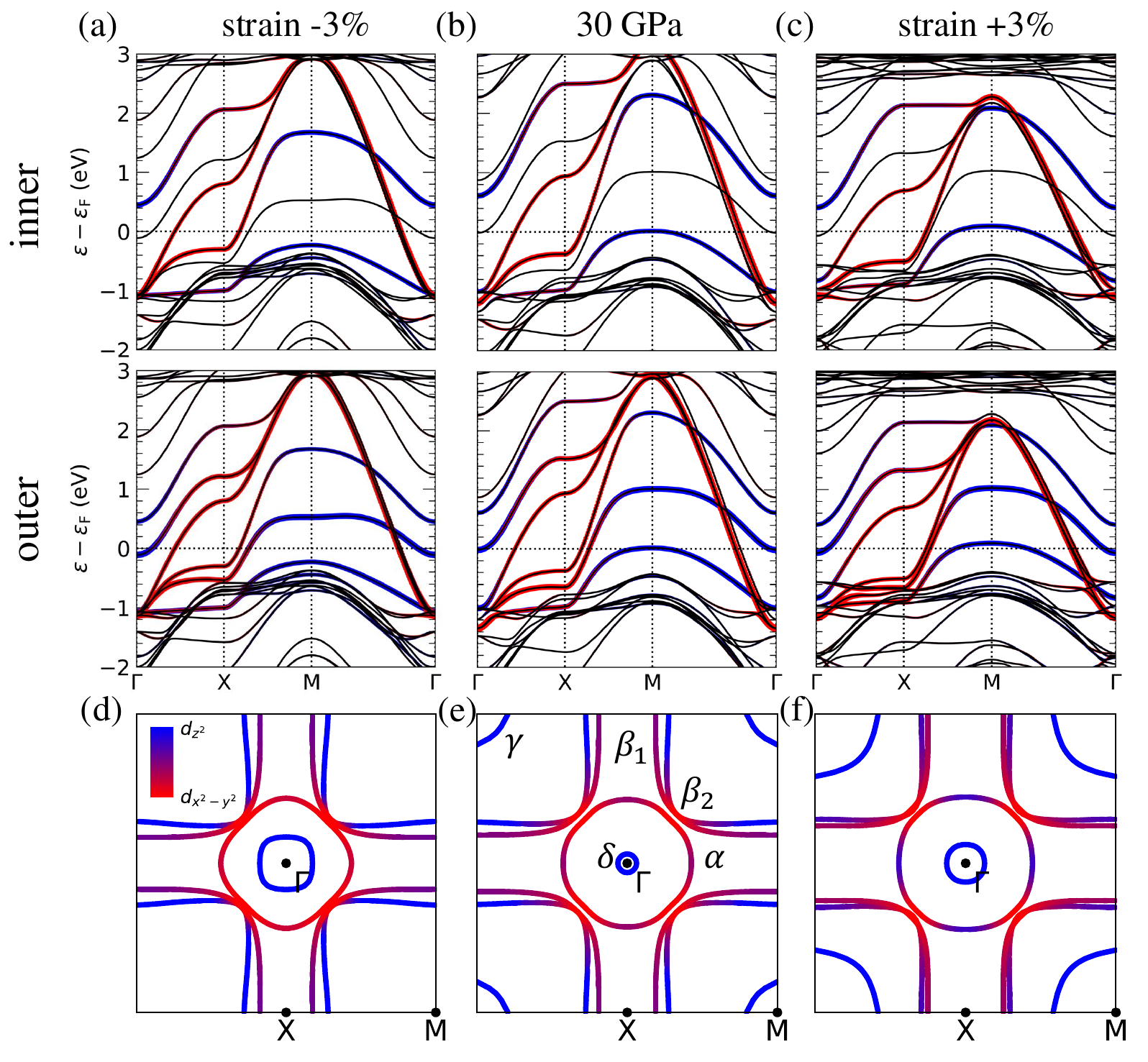}
    \caption{Band structure and Fermi surface of La$_4$Ni$_3$O$_{10}$ ($n=3$ RP) under different strains and under pressure with orbital characters highlighted in $I4/mmm$ symmetry: (a) and (d) panels for a $-3\%$ (compressive) strain, (b) and (e) panels for 30 GPa, and (c) and (f) panels for a $+3\%$ (tensile) strain. The band structure plots are atom-resolved for the inequivalent Ni atoms in the structure: inner and outer. }
    \label{fig4}
\end{figure}

Moving on to the effects of strain in the electronic structure of La$_4$Ni$_3$O$_{10}$, Fig. \ref {fig4} shows the band structure plots (resolved for each inequivalent Ni atom in the trilayer structure: inner and outer) and the corresponding Fermi surfaces under a 3\% compressive and tensile strain, in comparison to those obtained at 30 GPa. As shown in previous work \cite{Zhang2024, labollita2024trilayer}, the electronic structure of the trilayer RP at 30 GPa near the Fermi level is once again characterized by Ni-$e_g$ states hybridized with O-$p$ states. This is expected from formal valence considerations that in La$_4$Ni$_3$O$_{10}$ give rise to an average $d^{7.33}$ filling (corresponding to a Ni$^{2.67+}$ ion). As such, within each trilayer complex, four electrons need to be accommodated in the
 Ni-e$_g$ orbitals. As we have shown before \cite{labollita2024trilayer}, one should note that the $d_{z^2}$ orbitals are filled quite differently when comparing the inner-outer Ni atoms due to the formation of a bonding-nonbonding-antibonding molecular orbital complex with band splittings that are similar to those of La$_3$Ni$_2$O$_{7}$ (the odd symmetry nonbonding state does not involve the inner Ni site as can be clearly observed in the fat bands). In contrast, no relevant
distinction is obtained in the filling of the $d_{x^2-y^2}$ orbitals for the two inequivalent Ni atoms whose bandwidth is similar to that of the bilayer material at 30 GPa. Overall, five bands cross the Fermi level in La$_4$Ni$_3$O$_{10}$ under pressure, consequently contributing to the Fermi surface. Following an equivalent notation to that used for the bilayer nickelate, we label these bands as $\delta$, $\alpha$, $\beta_1$,  $\beta_2$, and $\gamma$. Similarly, the  $\gamma$ pocket is purely $d_{z^2}$ in character (the small $\delta$ pocket is also $d_{z^2}$ dominated) while the  $\alpha$, $\beta_1$, and $\beta_2$ sheets have a mixed $d_{z^2}$ and $d_{x^2-y^2}$ orbital character. The emergence of the $\gamma$ pocket is once again the main distinctive feature in the Fermi surface of La$_4$Ni$_3$O$_{10}$ under high pressure when compared to the ambient pressure results and its appearance has been linked to the presence of an $s^{\pm}$ leading pairing instability - the calculated pairing strength of the $s^{\pm}$ gap structure being smaller than that obtained for the bilayer RP \cite{Zhang2024}.

Similar trends under strain to those described in the previous section for the bilayer nickelate are obtained for La$_4$Ni$_3$O$_{10}$: under a 3$\%$ compressive strain the $d_{x^2-y^2}$ bandwidth is comparable to that obtained at 30 GPa but once again the $\gamma$ band of pure $d_{z^2}$ character vanishes from the Fermi surface as it is shifted down in energy (only the $\alpha$, $\beta_{1,2}$ and $\delta$ pockets are preserved under compressive strain giving rise to a more cuprate-like electronic structure). For a 3\% tensile strain, the $d_{z^2}$-$\gamma$ pockets are clearly recovered while the other Fermi surface sheets ($\alpha$, $\beta_{1,2}$ and $\delta$) are also kept. The same trends in the dominant hoppings described above for the bilayer compound are also obtained (see Fig. \ref{fig3}(b)). Hence, compressive strain seems to once again bring us closer to a cuprate-like electronic structure in La$_4$Ni$_3$O$_{10}$, while tensile strain mimics the appearance of the extra $\gamma$ pocket that also emerges under pressure (regardless of the increase in the out-of-plane Ni-O-Ni bond angle). In connection to this general finding, the suppression of octahedral tilts is likely not a sufficient condition for superconducting signatures to arise in RP nickelates. This conclusion can also be linked to the recent discovery of a single-layer$+$trilayer polymorph of La$_3$Ni$_2$O$_7$ where at ambient pressure no octahedral tilts are present in several structural refinements \cite{Chen2024poly, Wang2024longrange} (and there is no superconductivity at ambient pressure).

\subsection{Higher-order RPs}

After having analyzed the effects of strain in the two nickelate RPs wherein superconducting signatures under pressure have been reported in the bulk, we now proceed to analyze the effects of strain in higher-order ($n>3$) RPs. As mentioned above, any RP nickelate with $n>3$ is not thermodynamically stable in bulk form but they can be synthesized in thin film form using molecular beam epitaxy \cite{li2020epitaxial,Sun2021,Pan2022}. These higher-order RPs could, in principle, enable expanding the family of RP nickelate superconductors.

\begin{figure}
    \centering
    \includegraphics[width=\linewidth]{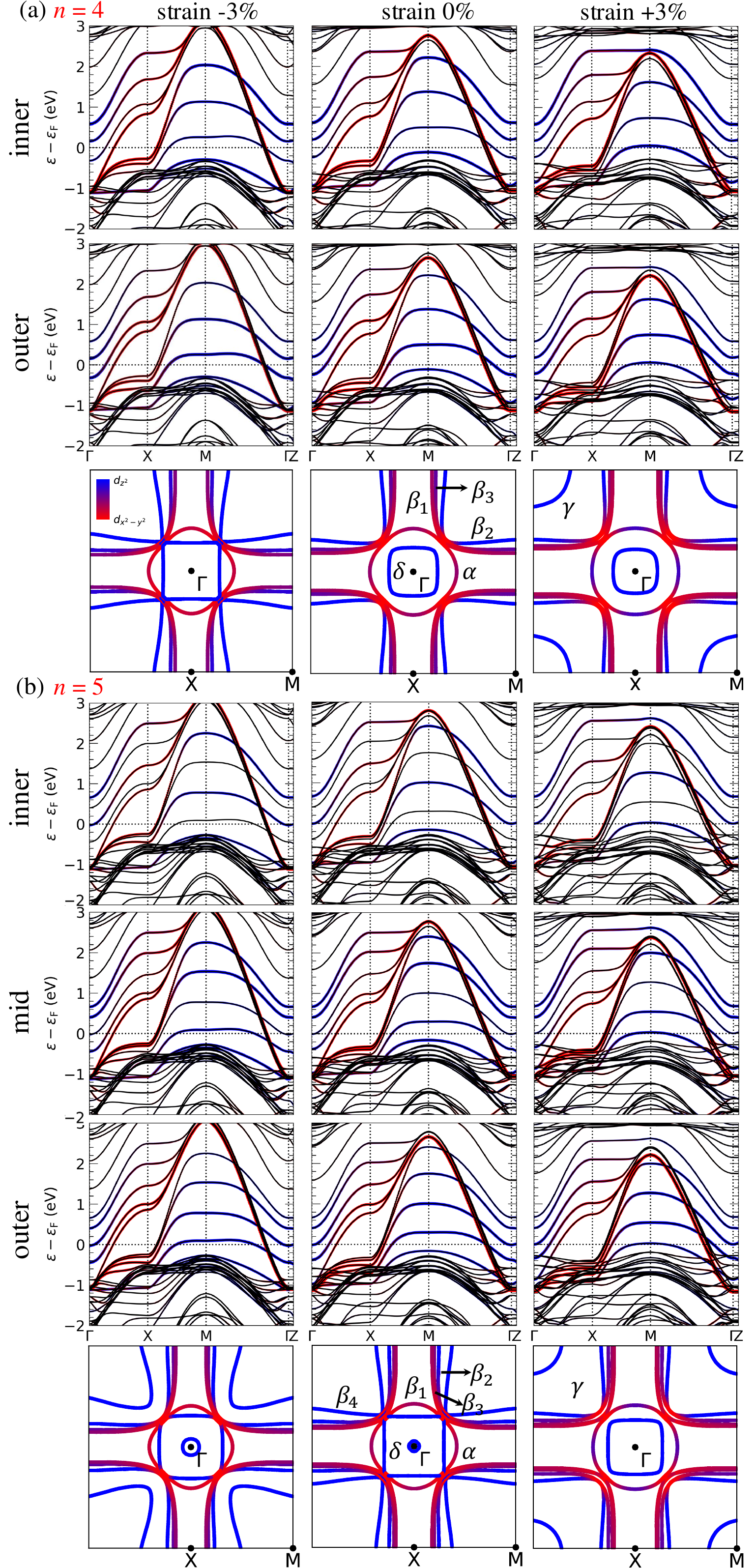}
    \caption{Band structure and Fermi surface of (a) La$_5$Ni$_4$O$_{13}$ ($n=4$ RP) and (b) La$_6$Ni$_5$O$_{16}$ ($n=5$ RP) under different strains with orbital characters highlighted in $I4/mmm$ symmetry. The band plots are atom-resolved for the different inequivalent Ni atoms in the structure: inner, outer for the $n=4$ RP;  inner, mid, and outer for the $n=5$ RP. All left panels are for a $-3\%$ (compressive) strain, middle panels for unstrained cases,  and right panels for a $+3\%$ (tensile) strain. }
    \label{fig5}
\end{figure}

Given that the structures of the higher-order RP nickelates have not been experimentally resolved, we analyze the trends in the electronic structure under strain for the $n=4$ and 5 RPs (La$_5$Ni$_4$O$_{13}$ and La$_6$Ni$_5$O$_{16}$) using the corresponding tetragonal phases as a starting point at ambient pressure. We note that for both La$_3$Ni$_2$O$_{7}$ and La$_4$Ni$_3$O$_{10}$ the electronic structure in the tetragonalized phases at ambient pressure is very close to that of the orthorhombic or monoclinic cells \cite{jung2022electronic, labollita2024electronic}. As applying large pressures to thin films will likely not be feasible, we simply compare the ambient pressure electronic structure of the $n=4$ and 5 RPs with that obtained under a 3\% compressive and tensile strain to extract some general trends. As expected from their average electron fillings ($d^{7.25}$ for La$_5$Ni$_4$O$_{13}$ and $d^{7.2}$ for La$_6$Ni$_5$O$_{16}$), only bands of $e_g$ character cross the Fermi level, like in the bilayer and trilayer counterparts. As in the $n=2$ and 3 compounds, the filling of the $d_{x^2-y^2}$ bands does not significantly differ among the inequivalent Ni atoms in the structure while the filling of the $d_{z^2}$ orbitals is markedly different due to the formation of molecular orbitals (whose eigenvalues and eigenvectors are reported in previous work \cite{jung2022electronic}). In the 4-layer RP La$_5$Ni$_4$O$_{13}$, there are 5 bands crossing the Fermi level at ambient pressure: two smaller electron-pockets centered at $\Gamma$ ($\delta$ of pure $d_{z^2}$ character and $\alpha$ of mixed $d_{z^2}$ and $d_{x^2-y^2}$ character) and three larger sheets also of mixed orbital character: $\beta_1$,  $\beta_2$, and $\beta_3$. The electronic structure of the 5-layer nickelate is very similar, with the addition of one extra $\beta$-like sheet as a consequence of the extra Ni. For none of the higher-order RPs there are $\gamma$-like pockets at ambient pressure. Under compressive strain, these pockets are also absent and the fermiology looks similar to that at ambient pressure, other than for the splittings between the $\beta$-like pockets being largely increased at X. In contrast, under tensile strain the $\gamma$ pocket of pure $d_{z^2}$ character (associated with the emergence of $s^{\pm}$-wave pairing in the bilayer and trilayer compounds) is recovered, further confirming the trends we reported in the previous subsections. The same trends in the dominant hoppings described above for the bilayer and trilayer compounds are also obtained for the higher order RPs as shown in Fig. \ref{figs5} of Appendix \ref{appendix:D}.

\section{summary and discussion}

By means of first-principles calculations, we have shown that strain is a viable way to decouple the structure and electronic-structure effects arising in La$_{n+1}$Ni$_n$O$_{3n+1}$ ($n= 2-5$) RP nickelates under pressure. Compressive strain increases the out-of-plane Ni-Ni distance as well as the apical Ni-O-Ni angle bringing it closer to 180$^\circ$. In terms of the electronic structure, it shifts the $d_{z^2}$ flat bands away from the Fermi level giving rise to a more cuprate-like electronic structure. Such a shift of the $d_{z^2}$ flat bands under compressive strain may favor $d$-wave superconductivity.  In contrast, tensile strain shortens the out-of-plane Ni-Ni distance and it decreases the apical Ni-O-Ni bond that further deviates from 180$^{\circ}$ with respect to the ambient pressure cases. In spite of the persistence of octahedral tilts, tensile strain recovers the flat $d_{z^2}$ bands crossing the Fermi level that have been associated with the emergence of $s^{\pm}$-wave pairing in the bilayer (La$_3$Ni$_2$O$_7$) and trilayer (La$_4$Ni$_3$O$_{10}$) RP nickelates. 
In this manner (biaxial) strain should be an interesting playground to pursue superconductivity in thin films of RP nickelates at ambient pressure as well as to investigate the competition between different pairing symmetries. Ultimately, it should also enable to pinpoint the structural and electronic structure characteristics necessary for superconductivity to emerge in this family of materials. 

\textit{Note added}. During the completion of this work, a
preprint  \cite{geisler2024} appeared reporting the structure and electronic structure of strained La$_3$Ni$_2$O$_7$ on LaAlO$_3$ and SrTiO$_3$. On SrTiO$_3$ (that provides $\sim$ 1\% tensile strain) the octahedral tilts are retained and the electronic structure resembles that obtained under pressure.

\section{acknowledgments}

We thank H. LaBollita, J Kapeghian, J. A. Mundy, and M. R. Norman for fruitful discussions. We acknowledge support from NSF Grant No. DMR-2323971, as well as the ASU Research Computing Center for HPC resources.

\onecolumngrid
\appendix

\section{\label{appendix:A} MLWFs for RP nickelates}
\begin{figure*}[h!]
    \centering
    \includegraphics[width=1.0\linewidth]{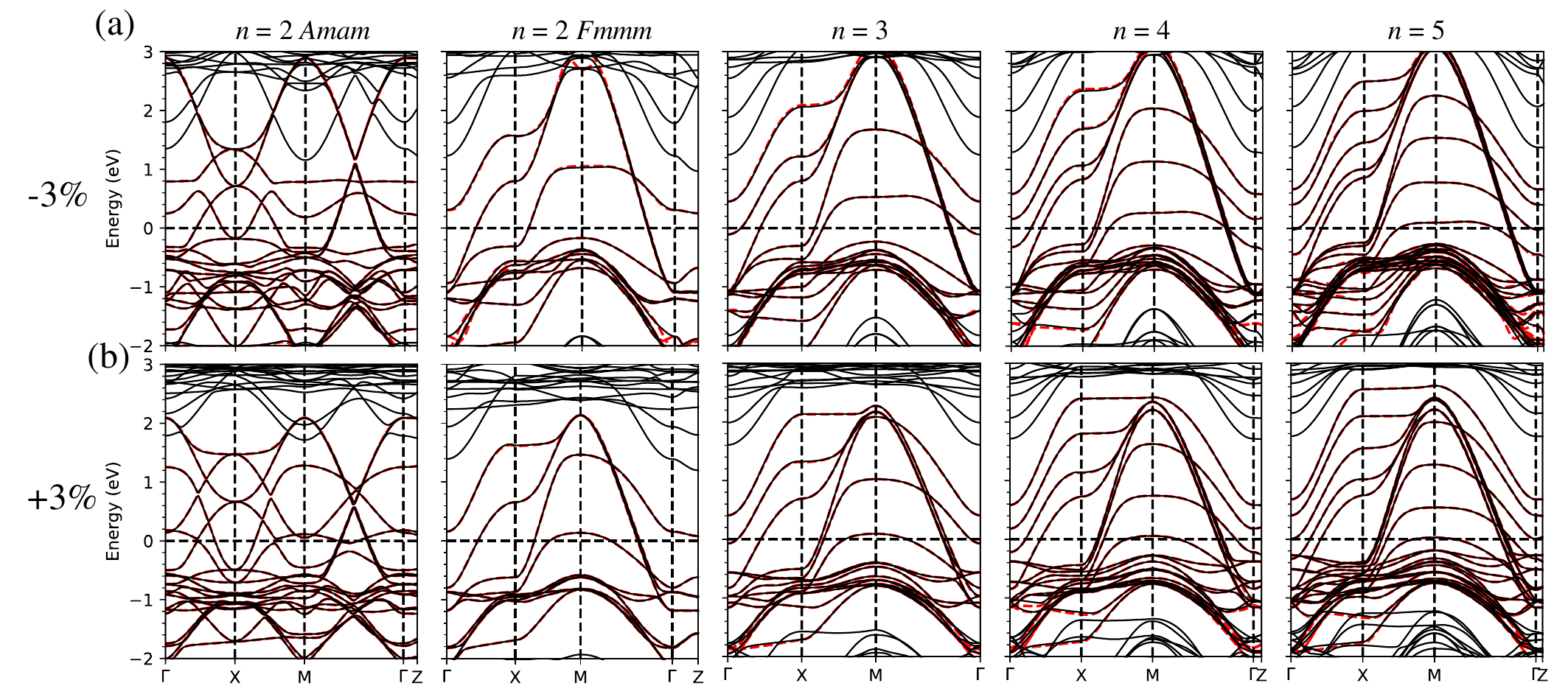}
    \caption{DFT band structures (black solid lines) and Ni-$d$ band wannierization (red dashed lines) as a function of strain for La$_{n+1}$Ni$_n$O$_{3n+1}$ ($n=2-5$) for a $3\%$ compressive strain (top panels, a) and for a $3\%$ tensile strain (bottom panels, b).}
    \label{figs1}
\end{figure*}

As described in the main text, we downfolded the DFT results onto the MLWFs to extract the dominant hopping parameters of the $n=2-5$ RP nickelates. 
In Fig. \ref{figs1} we show that the agreement between the band structures
obtained from Wannier function interpolation and those derived
from the DFT calculations is excellent, indicating
a faithful (though not unique) transformation to WFs.

\section{\label{appendix:E}Electronic structure of  strained bilayer nickelate La$_3$Ni$_2$O$_7$ in $Amam$ symmetry}

\begin{figure}[h!]
    \centering
    \includegraphics[width=0.5\linewidth]{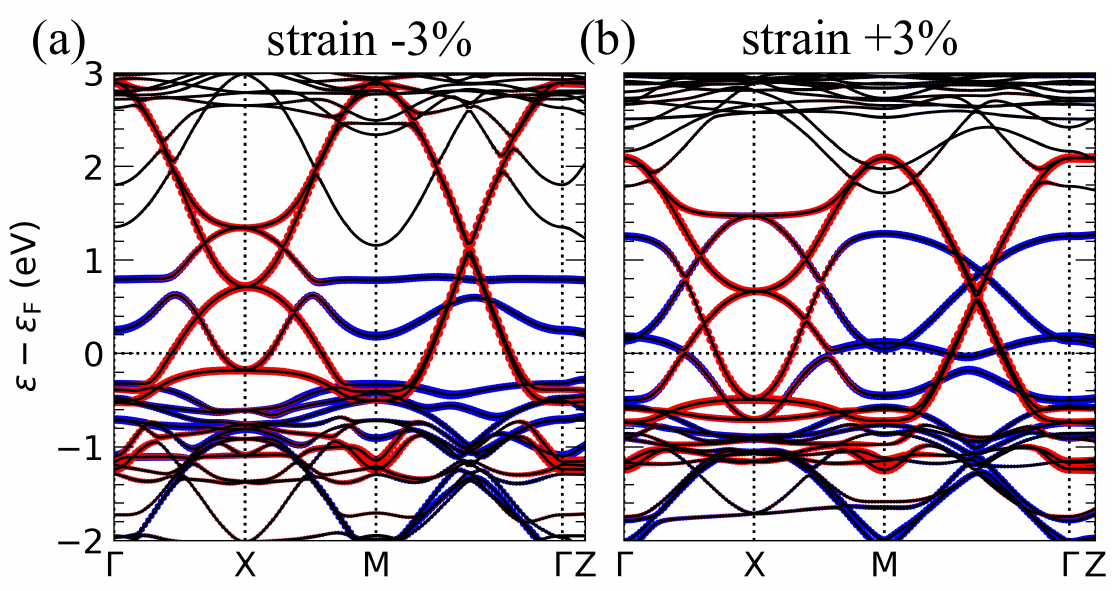}
    \caption{Band structure of La$_3$Ni$_2$O$_7$ ($n=2$ RP) under different strains with orbital characters highlighted in $Amam$ symmetry: (a) for a $-3\%$ (compressive) strain, (b) for a $+3\%$ (tensile) strain.}
    \label{figs2}
\end{figure}

The band structures for La$_3$Ni$_2$O$_7$ under a $\pm$3\% strain using an $Amam$ structure are shown in Fig.~\ref{figs2}. The same trends in terms of band shifts as those observed in $Fmmm$ symmetry (described in the main text) are retained: note in particular the bonding $d_{z^2}$ band being shifted below the Fermi level (at M) under compressive strain while it clearly crosses it under tensile strain.

\section{\label{appendix:C}Densities of states for RP nickelates}

\begin{figure*}
    \centering
    \includegraphics[width=1.0\linewidth]{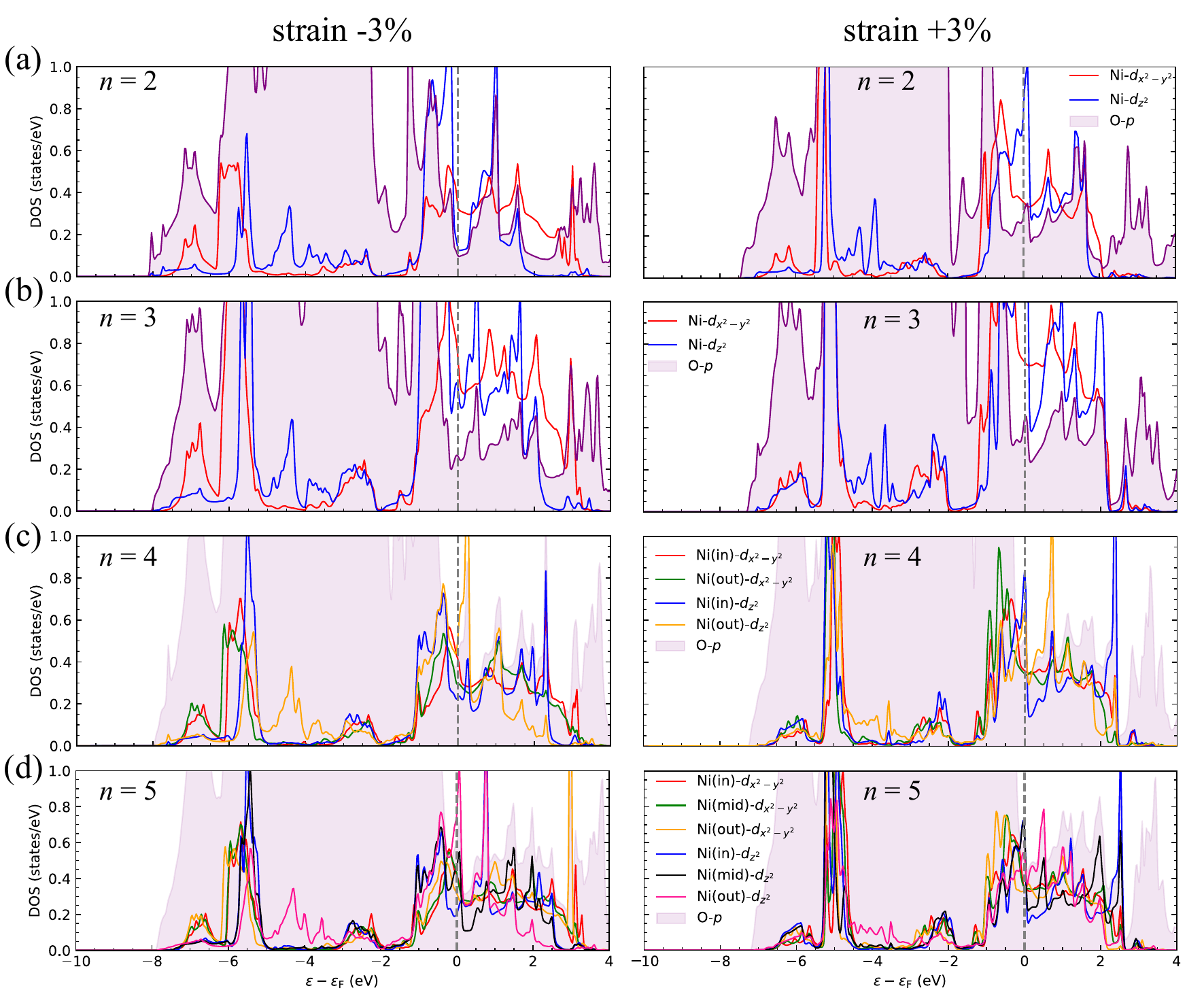}
    \caption{Orbital-resolved densities of states for Ni and O atoms in La$_3$Ni$_2$O$_7$ ($n=2$), La$_4$Ni$_3$O$_{10}$ ($n=3$), La$_5$Ni$_4$O$_{13}$ ($n=4$), and La$_6$Ni$_5$O$_{16}$ ($n=5$) under compressive ($-3\%$) and tensile ($+3\%$)  strains.}
    \label{figS4}
\end{figure*}

Figure \ref{figS4} shows the atom and orbital resolved DOSs for La$_{n+1}$Ni$_n$O$_{3n+1}$ ($n=2-5$) and their evolution under strain. The main changes described in the main text can be further scrutinized in the DOSs, in particular the changes in bandwidth upon applying strain as well as some of the relevant changes in orbital character.

\section{\label{appendix:B} Electronic structure of 1\% strained bilayer nickelate}

\begin{figure}[h!]
    \centering
    \includegraphics[width=0.5\linewidth]{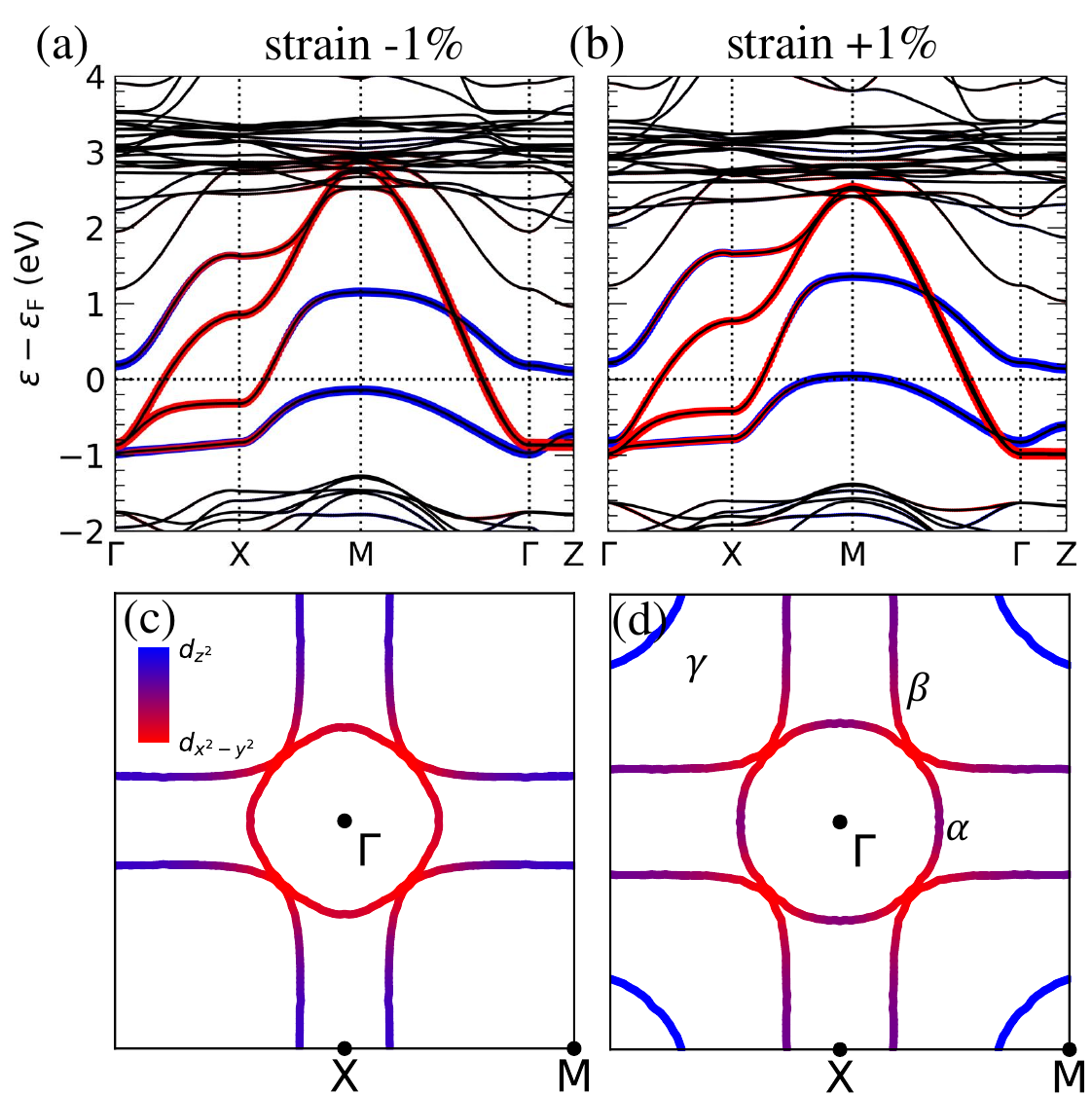}
    \caption{Band structure and Fermi surface of La$_3$Ni$_2$O$_7$ ($n=2$ RP) under different strains with orbital characters highlighted in $Fmmm$ symmetry: (a) and (c) panels for a $-1\%$ (compressive) strain, (b) and (d) panels for a $+1\%$ (tensile) strain.}
    \label{figs3}
\end{figure}

The band structures and Fermi surfaces for La$_3$Ni$_2$O$_7$ under a $\pm$1\% strain are shown in Fig.~\ref{figs3}. The same trends as those described in the main text are retained in that for compressive strain the $\gamma$ pockets are absent at the zone corners but they appear under tensile strain. As mentioned above, the size of these $\gamma$ pockets for a 1\% tensile strain is actually closer to that obtained at 30 GPa. However, as the lattice constants obtained for a 1\% strain are outside of the superconducting regime in La$_3$Ni$_2$O$_7$, we focus on the main text on larger levels of strain.

\section{\label{appendix:D}Hopping parameters for $n=4-5$ RPs under strain}

\begin{figure}[h!]
    \centering
    \includegraphics[width=0.5\linewidth]{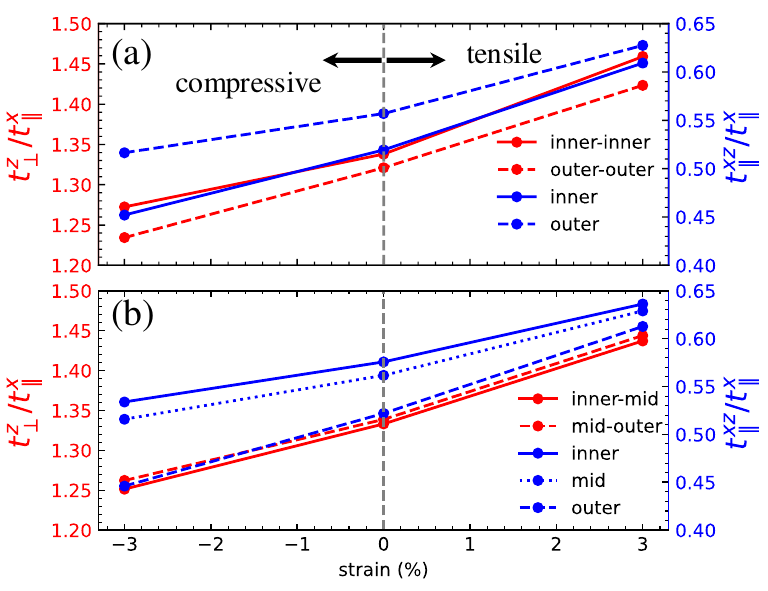}    \caption{Evolution of the dominant hopping ratios for (a) La$_5$Ni$_4$O$_{13}$ ($n=4$ RP) and (b) La$_6$Ni$_5$O$_{16}$ ($n=5$ RP) as a function of strain.}
    \label{figs5}
\end{figure}

Figure \ref{figs5} shows the evolution of the dominant hopping ratios for the 4-layer and 5-layer RP nickelates under strain. The same trends as those described in the main text for the bilayer and trilayer nickelates are observed. 

\clearpage
\twocolumngrid
\bibliography{stress}

\end{document}